\documentclass[manuscript,screen]{acmart}
\AtBeginDocument{%
  \providecommand\BibTeX{{%
\normalfont B\kern-0.5em{\scshape i\kern-0.25em b}\kern-0.8em\TeX}}}

\setcopyright{rightsretained}
\copyrightyear{2023}
\acmYear{2023}
\acmDOI{}

\acmConference[CCAI 2023]{CHI 2023 Workshop on Child-centred AI Design: Definition, Operation and Considerations}{April 23, 2023}{Hamburg, Germany}
\acmBooktitle{CHI 2023 Workshop on Child-centred AI Design: Definition, Operation and Considerations, April 23, 2023, Hamburg, Germany}

\begin{document}

\title{Adapt a Generic Human-Centered AI Design Framework in Children’s Context}

\author{Zhibin Zhou}
\affiliation{%
  \institution{School of Design, The Hong Kong Polytechnic University}
  \city{Hong Kong}
  \country{Hong Kong, China}}
\email{zhibin.zhou@polyu.edu.hk}

\author{Junnan Yu}
\affiliation{%
  \institution{School of Design, The Hong Kong Polytechnic University}
  \city{Hong Kong}
  \country{Hong Kong, China}}
\email{junnan.yu@polyu.edu.hk}


\begin{abstract}
  Through systematically analyzing the literature on designing AI-based technologies, we extracted design implications and synthesized them into a generic human-centered design framework for AI technologies to better support human needs and mitigate their concerns (Figure 1) \cite{sun23}. As the figure shows, the framework consists of four dimensions: Machine Learning, Stakeholders, Context, and UX Values. Each dimension further includes components and characteristics impacting user experiences. Although not explicitly situated in children’s context, the framework has revealed important design and research directions for designing AI-based technologies for children, to name a few in different dimensions:

  \begin{itemize}
  \item \textit{Data privacy and security (Machine Learning):} Appropriate measures are needed to ensure children’s data is used ethically and responsibly. Also, we need to raise children’s and caregivers’ awareness of what data AI systems will use in which ways.  
  \item \textit{Ethical considerations (Stakeholders):} There are many ethical considerations when developing AI systems for children, especially potential harmful consequences, such as unfair algorithms, reinforced biases, and resulting abuses. We will need a systematic exploration of ethical considerations for children’s AI technologies, including existing ones for general technologies and new ones unique to AI systems. 
  \item \textit{Diverse abilities and needs (UX Values):} Even children of a similar age have different developmental and cognitive abilities, not to mention children of different ages. How can technology be designed to accommodate these differences? There is still limited data on how children interact with various AI applications, making it challenging to develop and test AI systems appropriate for their needs and abilities.
  \item \textit{Social networks (Context):} Children’s technology use is typically mediated and regulated by their caregivers, especially parents. Accordingly, how caregivers’ expectations, concerns, and interventions regarding their children’s use of AI technologies should be comprehensively investigated and considered in the design.
\end{itemize}

When adapting the framework to children’s context, understanding their specific needs, behaviors, experiences, and social environments is needed. Therefore, we are working on projects to explore tailored design considerations for children, such as through investigating children’s use of existing AI-based toys and learning technologies. By participating in the ACM CHI 2023 Workshop on \textit{“Child-Centred AI Design: Definition, Operation, and Considerations,”} we hope to learn more about how other researchers in this field approach designing child-centered AI technologies, exchange ideas on the research landscape of children and AI, and explore the possibility to develop a practical child-centered design framework of AI technologies for technology designers and developers. 

\end{abstract}




\begin{teaserfigure}
  \includegraphics[width=\textwidth]{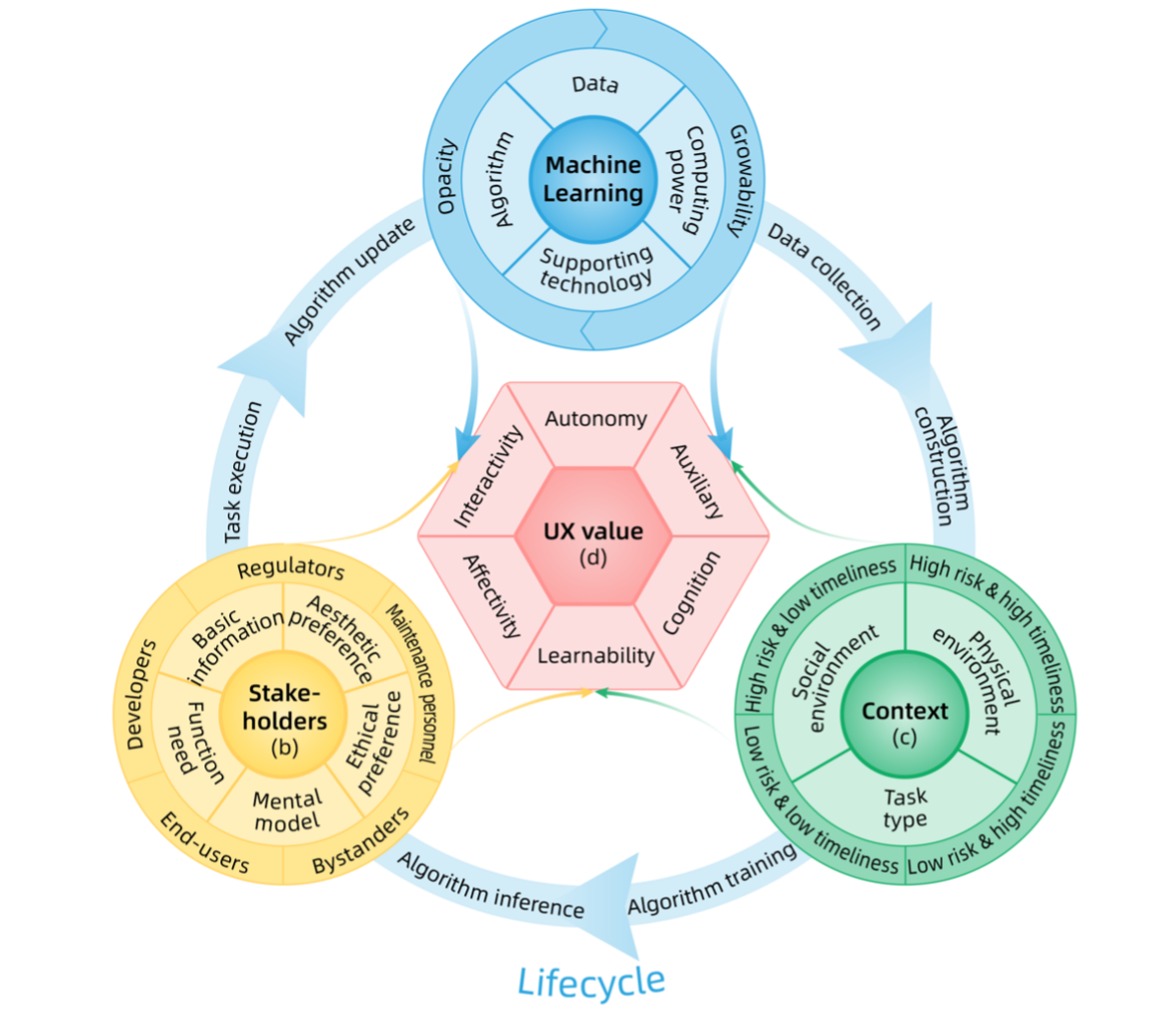}
  \caption{The generic human-centered UX design framework for AI technologies \cite{sun23}}
  \label{fig:teaser}
\end{teaserfigure}

\maketitle


\bibliographystyle{ACM-Reference-Format}
\bibliography{sample-base}

\end{document}